\documentclass[twocolumn,times]{aastex701}
\hypersetup{linkcolor=blue,citecolor=blue,filecolor=cyan,urlcolor=blue}

\usepackage{mhchem}
\usepackage{mathptmx}
\newcommand{\kcalmol}{kcal mol$^{-1}$}
\newcommand{\rev}[1]{#1}

\shorttitle{Carbon chemisorption on solid CO}
\received{\today}
\revised{\today}
\accepted{\today}
\submitjournal{ApJ}

\begin{document}

\title{Formation of unsaturated carbon chains through carbon chemisorption on solid CO}

\author[orcid=0000-0001-9669-1288]{Masashi Tsuge}
\affiliation{Institute of Low Temperature Science, Hokkaido University, Sapporo 060-0819, Kita-19 Nishi-8, Kita-ku, Sapporo 060-0819, Japan}
\email[show]{tsuge@lowtem.hokudai.ac.jp}  

\author[orcid=0000-0001-8803-8684]{Germán Molpeceres} 
\affiliation{Departamento de Astrofísica Molecular, Instituto de Física Fundamental (IFF-CSIC), C/ Serrano 121, 123, 113B E-28006 Madrid, Spain}
\email[show]{german.molpeceres@iff.csic.es}

\author[orcid=0009-0002-3208-3296]{Ryota Ichimura}
\affiliation{Division of Science, National Astronomical Observatory of Japan, 2-21-1 Osawa, Mitaka, 181-8588, Japan}
\affiliation{Department of Astronomical Science, The Graduate University for Advanced Studies, SOKENDAI, 2-21-1 Osawa, Mitaka, 181-8588, Japan}
\email{fakeemail3@google.com}

\author[orcid=0000-0002-7058-7682]{Hideko Nomura}
\affiliation{Division of Science, National Astronomical Observatory of Japan, 2-21-1 Osawa, Mitaka, 181-8588, Japan}
\email{fakeemail4@google.com}

\author[orcid=0000-0002-2026-8157]{Kenji Furuya}
\affiliation{RIKEN Pioneering Research Institute, 2-1 Hirosawa, Wako,  351-0198, Japan}
\email{fakeemail5@google.com}

\author[orcid=0000-0001-8408-2872]{Naoki Watanabe}
\affiliation{Institute of Low Temperature Science, Hokkaido University, Sapporo 060-0819, Kita-19 Nishi-8, Kita-ku, Sapporo 060-0819, Japan}
\email{fakeemail5@google.com}

\begin{abstract}

The interaction of carbon atoms with solid carbon monoxide (CO) is a fundamental process in astrochemistry, influencing the formation of complex organic molecules in interstellar environments. This study investigates the adsorption and reaction mechanisms of carbon atoms on solid CO under cryogenic conditions, employing a combination of experimental techniques, including the combination of photostimulated desorption and resonance-enhanced multiphoton ionization (PSD-REMPI) and infrared spectroscopy, alongside quantum chemical calculations. The results reveal the formation of oxygenated carbon chains, such as \ce{CCO}, \ce{C3O2}, and \ce{C5O2}, as well as \ce{CO2}. The findings highlight the role of chemisorption and subsequent reactions in driving molecular complexity on solid CO, with implications for the chemical evolution of interstellar ices and the potential formation of prebiotic molecules.

\end{abstract}



\section{Introduction}

Carbon monoxide (CO) holds a unique position among interstellar molecules. It is not only the second most abundant molecule in the interstellar medium (ISM) after molecular hydrogen but also serves as a key precursor for the formation of interstellar complex organic molecules (COMs) \citep{Herbst2009}. A prominent example is the synthesis of interstellar methanol (\ce{CH3OH}), one of the most iconic organic molecules in space, which occurs via the hydrogenation of CO (\ce{CO + 4H -> CH3OH}) on ice-covered submicron-sized dust grains \citep{Watanabe2002,tsuge_radical_2023}. These icy grains form under the extremely low temperatures of dark molecular clouds (approximately 10~K), with their chemical composition evolving over time to represent different ice epochs \citep{Boogert2015}. The morphology of these ices during these epochs remains uncertain. Some studies propose that interstellar ices are stratified in an onion-like structure \citep{Boogert2015}, where an apolar layer forms atop a polar one, primarily composed of water, due to the differing timescales for polar and apolar material formation. Other studies suggest that different ice phases coexist without fully wetting the grain’s surface area \citep{2020PhRvL.124v1103P,Kouchi2021,kouchi_transmission_2021}. Regardless of the ice morphology, CO is expected to be the second most abundant constituent of interstellar ices and partially exist as CO solid. Therefore, interaction of adsorbates with CO solid plays a critical role in various chemical processes at the ice-gas interface.

To simulate energetic processes on interstellar icy grains, the irradiation of solid CO with electrons, protons, ultraviolet (UV) photons, heavy ions, and X-rays, has been extensively investigated in laboratory settings \citep{jamieson_understanding_2006,forstel_pentacarbon_2016,trottier_carbonchain_2004,1996A&A...312..289G,sie_key_2022, seperuelo_duarte_laboratory_2010,ciaravella_chemical_2016}. Under high-energy irradiation, the observed chemical processes are primarily attributed to the dissociation of CO molecules: \ce{CO(X^{1}\Sigma^{+}) -> C(^{3}P/^{1}D) + O(^{3}P/^{1}D)}. The resulting \ce{C(^{3}P/^{1}D)} atoms (hereafter designated as $^3$C and $^1$C) readily react with neighboring CO molecules to form \ce{^3CCO}, the simplest carbon chain intermediate. This species can subsequently undergo further reactions, initiating an oligomerization process. Specifically, \ce{C_{n}O} species (where $n \geq 2$) can react with either atomic carbon or CO to yield \ce{C_{n+1}O} or \ce{C_{n+1}O2}, respectively \citep{jamieson_understanding_2006}. In the case of UV irradiation with photon energies below 11~eV (the dissociation energy of CO), photodesorption is the dominant process \citep{sie_key_2022}. However, the formation of \ce{CO2} has also been observed. This phenomenon has been attributed to reactions involving excited CO (\ce{CO^*}), such as \ce{CO^* + CO -> CO2 + C} or, more generally, \ce{CO^* + C_{n}O -> CO2 + C_{n}} \citep{devine_spin-forbidden_2022}.

A completely different paradigm for chemistry on ices, particularly solid \ce{H2O} and \ce{CO}, involves the continuous accretion of atoms and small molecular species onto ice grains, a process characteristic of dark interstellar clouds. In these environments, adsorption of material occurs more frequently than photodissociation on icy due to the low permeability of photons in dark clouds. Some of the most common adsorbates landing on interstellar \ce{CO} include atoms such as H, C, O, and N. The traditional view of surface astrochemistry assumes that these adsorbates remain physisorbed on the ice surface, where they can diffuse and react with other species following the Langmuir–Hinshelwood mechanism \citep{tielens_model_1982}. In this context, hydrogenation reactions are considered the most significant pathway for increasing chemical complexity on icy grains, owing to the high mobility of hydrogen atoms \citep{Hama2012,Kimura2018} and their ability to quantum tunnel through kinetic barriers. Recent refinements of the diffusive mechanism include non-thermal scenarios, where newly formed molecules can transiently diffuse \citep{jin_formation_2020}. The behavior of fundamental species other than H atom, such as C, O, and N atoms, has not been studied in as much detail. Among these, the C atom is particularly important because its reactions with other species can induce skeletal evolution of molecules, such as C–C bond formation \citep{tsuge_surface_2023}. Especially intriguing is the duality exhibited by C atoms, as demonstrated in their behavior on water ice. In this case, C atoms can either remain physisorbed, facilitating molecular evolution, or chemisorb through an Eley–Rideal mechanism, depending on the adsorption binding site. This duality opens entirely different chemical pathways for the formation of complex molecules  \citep{tsuge_radical_2023,tsuge_surface_2023,ferrero_formation_2024,Molpeceres2021c}.

The reaction of C atoms with \ce{CO} in a low-temperature solid has been recently studied both experimentally \citep{fedoseev_hydrogenation_2022}, and theoretically \citep{ferrero_formation_2023}. \rev{\citet{fedoseev_hydrogenation_2022}} performed experiments where H/C/\ce{CO}/\ce{H2O} were simultaneously deposited at cryogenic temperatures. Under these conditions, they observed the formation of ketene (\ce{H2CCO}), leading them to conclude that the \ce{CCO} molecule formed by the reaction \ce{^{3}C + ^{1}CO -> ^{3}CCO} was subsequently hydrogenated to produce \ce{H2CCO}. Although this pathway is very plausible under their experimental conditions, direct evidence for the formation of \ce{^{3}CCO} was not presented. \rev{\citet{krasnokutski_pathway_2022} performed co-deposition of C/CO/\ce{NH3} and observed infrared features attributable to molecules with a C=C=O moiety.} However, no definitive assignments were given in their paper. The formation of \ce{CCO} and related products was later investigated computationally \citep{ferrero_formation_2023}. In this work, we employ our novel experimental setup to detect surface adatoms at cryogenic temperatures \citep{tsuge_surface_2023,tsuge_methane_2024}, posing the question of whether co-deposition experiments fully reflect the chemistry occurring under the conditions of cold dark clouds or even protoplanetary discs, which are characterized by a slow accretion of molecular material into solid bodies. In co-deposition experiments, rapid deposition of reactants facilitates immediate reactions. Recently, \rev{\citet{tsuge_methane_2024}} demonstrated that methane (\ce{CH4}) production yield on the surface of amorphous solid water (ASW) is significantly different depending on the deposition method; co-deposition of C and H atoms yielded almost 100\% conversion of C atoms into methane, whereas sequential deposition (C atom deposition followed by H atom deposition) resulted in less than 10\% conversion. If the same paradigm-changing conclusions found for water ice are repeated in solid \ce{CO}, it would change our view on the formation of unsaturated carbon chains, which is traditionally attributed to gas-phase chemistry \citep{sakai_warm_2013}. To shed light on this conundrum, we undertook the task of simulating the adsorption of carbon atoms on solid \ce{CO}. To succeed in this task, we applied the combination of photostimulated desorption and resonance-enhanced multiphoton ionization methods (PSD-REMPI) for the \textit{in situ} detection of C atoms on solid \ce{CO}, in combination with tailored quantum chemical calculations of reactivity. The reaction products in the sequential deposition (C atom deposition on solid \ce{CO}) and co-deposition experiments were studied by infrared (IR) spectroscopy, including isotopic labeling. More details on Methodology are given in Section \ref{sec:experimental}.

\section{Methods} \label{sec:experimental}

\subsection{Experimental Apparatus} 

All experiments were performed using the apparatus named RASCAL (Reaction Apparatus for Surface Characteristic Analysis at Low-temperature) at the Institute of Low Temperature Science, Hokkaido University. The details of RASCAL have been provided in previous works \citep{Hama2012, Watanabe2010}. RASCAL consists of an ultrahigh vacuum main chamber (with a base pressure of $\sim$10$^{-8}$ Pa) and a differentially pumped one for the C-atom source \rev{chamber}. The aluminum (Al) substrate was positioned at the center of the main chamber and can be cooled to 5.5 K using a closed-cycle He refrigerator (RDK-415E, Sumitomo Heavy Industries). Temperature of the substrate was controlled using a silicon-diode temperature sensor (DT-670, Lake Shore), ceramic heaters (MC1020, Sakaguchi E. H VOC), and a temperature controller (Model 335, Lake Shore).

A commercial C-atom source (SUKO-A, MBE-Komponenten) was used to provide exclusively C[1] in its electronic ground state; hence, no C atom clusters such as \ce{C3} are emitted from the source. The most significant contaminant from the source was CO, which is presumably in the electronic ground state with ro-vibrational excitations considering the energy levels and source temperature above 2000 K. The flux of the C atom was controlled by adjusting the temperature of the filament, which comprised a thin Ta tube filled with carbon powder. The flux was determined by co-depositing C atoms and excess \ce{O2}, where one \ce{O3} molecule, which can be quantified by IR spectroscopy, is produced from one C atom (\ce{C + O2 -> CO + O}; \ce{O + O2 -> O3}); i.e., the C-atom flux was calculated by dividing the column density of \ce{O3} by deposition time.

\subsection{Preparation of amorphous solid CO}

The amorphous solid CO samples were produced via background vapor deposition of CO molecules at a substrate temperature of 10 K. The deposition rate was approximately 1 monolayer (ML) min$^{-1}$, corresponding to the flux of $\sim$2$\times$10$^{13}$ atoms cm$^{-2}$ s$^{-1}$. This flux is slightly above the critical flux for the formation of amorphous CO at 10 K \citep{Kouchi2021}. In the PSD-REMPI measurements, CO solid with 5 MLs thickness were initially prepared and after each measurement additional 2-3 MLs were deposited to refresh the surface condition; this procedure was repeated for five to six measurements and after that the sample was sublimated.

\subsection{C-atom detection by PSD-REMPI}

The PSD-REMPI method has been successfully applied to detect trace reactive species such as H atoms and OH radicals on the surface of dust grain analogues \citep{Kuwahata2015a,miyazaki_photostimulated_2020, sie_photodesorption_2024}. Carbon atoms on solid CO were desorbed by weak nanosecond laser pulse at 532 nm (10 Hz repetition rate) with the typical pulse energy of 80 $\mu$J. The C atoms desorbed from the surface were then ionized by the REMPI laser that was focused approximately 1 mm above the Al substrate with the timing that achieves the maximum PSD-REMPI signals. At the focal point of the REMPI laser, C atoms were ionized using the (2+1)REMPI scheme via the 2s$^{2}$2p3p($^{3}$D$_{3}$) $\leftarrow$ 2s$^{2}$2p$^{2}$ ($^{3}$P$_{2}$) transition \citep{fassett_laser_1985}. Laser radiation in the wavelength range 286.8--286.9 nm with a pulse energy of approximately 1 mJ was provided from an \ce{Nd$^{3+}$}:YAG laser-pumped dye laser, with subsequent frequency doubling with a potassium dihydrogen phosphate crystal. The ionized C atom (\ce{C+}) was then detected by a linear type time-of-flight mass spectrometer equipped with a micro-channel plate detector.

\subsection{Infrared measurements} \label{sec:exp:ir}

The IR spectra of the samples on the Al substrate were measured using reflection-absorption IR spectrometry with a Fourier-transform infrared spectrometer (Spectrum One, Perkin Elmer) equipped with a KBr beam splitter and an Hg--Cd--Te detector. Spectra in a region of 700--4000 cm$^{-1}$ were collected at a resolution of 4 cm$^{-1}$ after averaging 200 or 400 scans.
The surface number density (column density) of species X, denoted as $\left[\rm X\right]$ in molecules cm$^{-2}$, was estimated using the following equation:

\begin{equation} \label{eq:exp}
  \left[X\right] = \dfrac{cos\left(\theta\right)\ln10}{2B}\int_{\nu_{1}}^{\nu_{2}}A\left(\nu\right)d\nu
\end{equation}
where $\theta$, $B$, and $A\left(\nu\right)$ represent the incident angle of the IR beam with respect to the Al substrate (83$^{\circ}$
), integrated absorption coefficient (in the unit of cm molecule$^{-1}$), and absorbance at a given wavenumber, respectively. The $B$ values for \ce{CO2} ($\nu_{3}$, $B$ = 1.0 $\times$ 10$^{-16}$), CCO ($\nu_{1}$, $B$ = 2.3 $\times$ 10$^{-17}$), \ce{C3O2} ($\nu_{3}$, $B$ = 4.9 $\times$ 10$^{-16}$), and \ce{C5O2} ($\nu_{4}$, $B$ = 1.1 $\times$ 10$^{-17}$) were calculated at the B3LYP/cc-pVTZ level of theory. These calculated values agreed very well with those provided by \rev{\citet{jamieson_understanding_2006}} based on their own B3LYP/6-311G** calculations. Although experimental $B$ values have been reported for solid \ce{CO2} and \ce{C3O2}, we used calculated value for consistency among all the species. \rev{value for \ce{CO2} agreed well with the reported experimental value of 1.8$\times$10$^{-16}$ \citep{gerakines_first_2015}. On the other hand, the $B$ value calculated for \ce{C3O2} is several times greater than the experimental value of 1.3$\times$10$^{-16}$ \citep{gerakines_carbon_2001};} this difference might originate from the fact that calculation was done for isolated molecule and experimental value is for the pure solid of \ce{C3O2}. We thus indicate that the \ce{C3O2} number density derived using calculated $B$ value could be overestimated.

\subsection{Quantum Chemical Calculations} \label{sec:theory}

To mimic the conditions of the sequential deposition experiments we studied the reaction of $^{3}$C with solid CO under high dilution conditions. To achieve that, we used a cluster model mimicking the local environment of a CO, extracted from our previous work \citep{molpeceres_cracking_2023}, containing 33 CO molecules. The level of theory throughout the calculations is also equivalent to that of that work (i.e. DLPNO-CCSD(T)/CBS//MN15(D3BJ)/6-31+G(d,p)) \citep{riplinger_sparse_2016, guo_communication_2018, zhong_uniformly_2008, neese_revisiting_2011, C6SC00705H, Grimme2010, Grimme2011, clark_efficient_1983}, where CBS stands for complete basis set (extrapolation), that we achieved with a two-point formula and DLPNO-CCSD calculations using the cc-pVDZ and cc-pVTZ basis-sets \citep{Woon1994}. The initial cluster structure is optimized at the MN15(D3BJ)/6-31+G(d,p) on top of which DLPNO-CCSD(T)/CBS single point calculations are performed. The reaction of C atoms with solid CO is simulated placing the atoms in random positions of the cluster and performing a force minimization. The chemisorption energy of the C atom is calculated as the difference between the energy of the cluster with the C atom and the energy of the cluster and carbon atoms separated. A similar procedure is later carried out, but now placing a \ce{C2} molecule close to a \ce{CCO} chemisorbate, in order to investigate \ce{C5O2} formation. The evolution of the chemisorbates reacting with a neighbor CO molecule is computed by performing potential energy scans along the reaction coordinate, and optimizing the local maxima found during the scan. Activation energies are then computed as the difference between the energy at the transition state and of the chemisorbate. All energies reported in this work include harmonic vibrational zero point energy corrections. The uncertainties in the energies correspond to the standard deviation of the energetic magnitudes in the three sampled binding sites. Geometry optimizations and frequency calculations were performed using the \textsc{Gaussian 16} program \citep{g16}. The DLPNO-CCSD(T) calculations were performed using the \textsc{Orca} 6.0.0 \citep{Neese2012, Neese2020} program.

\section{Experimental results} \label{sec:main:experimental}

\subsection{Characterization of C atoms on solid CO} \label{sec:experimental:rempi}

\begin{figure}
  \centering
  \includegraphics[width=\linewidth]{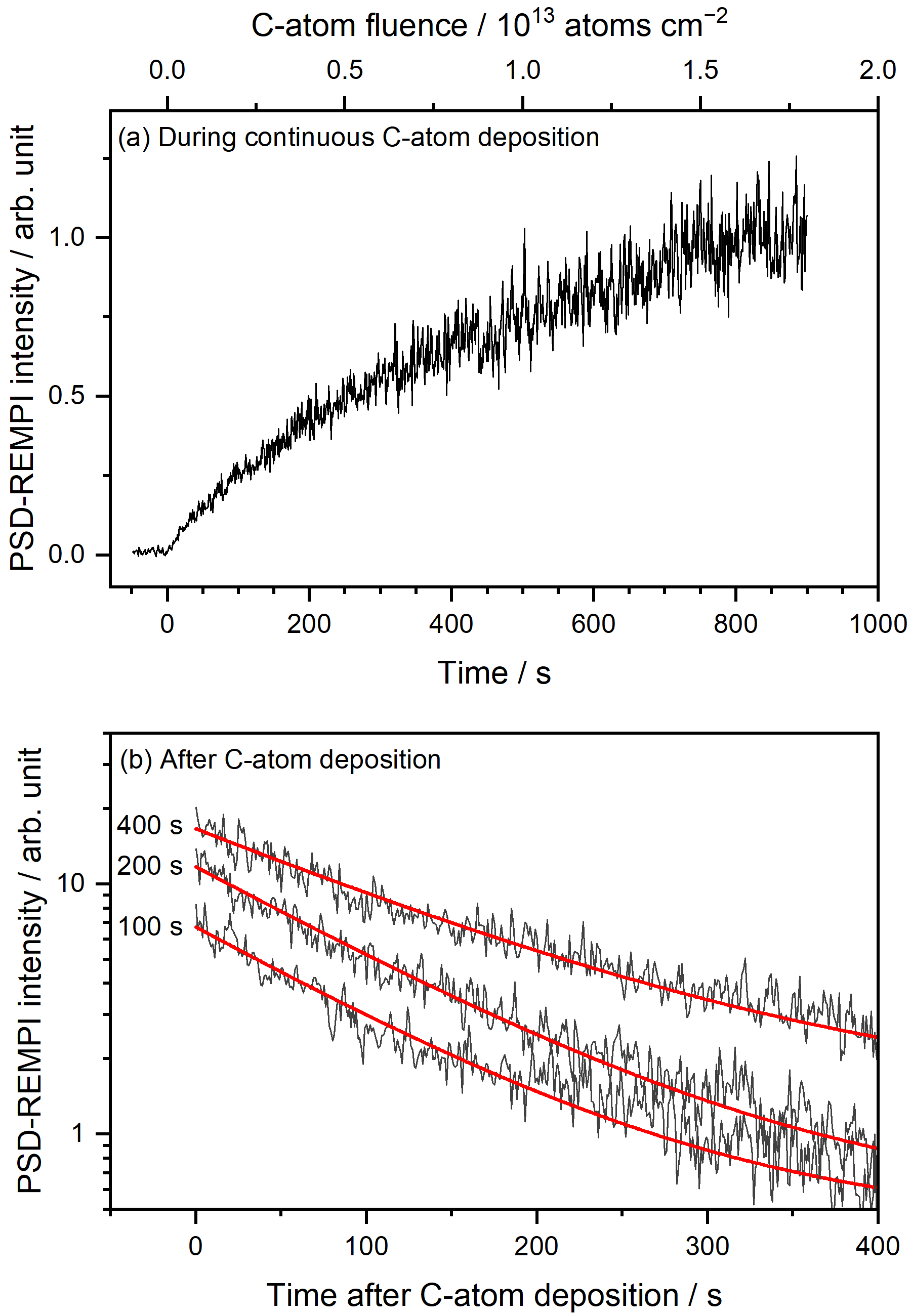}
  \caption{In situ C-atom detection by PSD-REMPI. (a) The PSD-REMPI intensity of C-atoms during continuous C-atom deposition on solid CO at 10 K with the C-atom flux of 2$\times$10$^{10}$ atoms cm$^{-2}$ s$^{-1}$. (b) The decay of PSD-REMPI intensity after terminating C-atom deposition; after 100 s deposition (bottom), 200 s (middle), and 400 s (top) deposition, corresponding to the C-atom fluences of 2$\times$10$^{12}$, 4$\times$10$^{12}$, 8$\times$10$^{12}$ atoms cm$^{-2}$, respectively. Black lines are experimental data and red lines represent the fitting result according to Equation \ref{eq:exp2}.}
  \label{fig:exp1}
\end{figure}

In panel~(a) of Figure~\ref{fig:exp1}, we show the PSD-REMPI intensity of $^3$C atoms, which is proportional to the surface number density, as a function of C-atom deposition time. The increasing rate of signal intensity gradually decelerates towards saturation. This behavior is similar to that observed for C atoms on ASW \citep{tsuge_surface_2023} and indicates physical processes that reduce the number of detectable carbon atoms. This behavior on ASW was attributed to the conversion of C atoms from a physisorbed state to a chemisorbed one \citep{tsuge_surface_2023}. By analogy, it is anticipated that the formation of the \ce{^{3}CCO} molecule from C atoms physisorbed on solid \ce{CO} might be the primary loss process, as predicted by theoretical studies \citep{ferrero_formation_2023}.\rev{The physisorbed states in our experiment cannot be clearly isolated contrary to our works on \ce{H2O} ice \citep{tsuge_surface_2023,tsuge_methane_2024}, because they transiently convert to chemisorbed ones, which is the unique state at the end of the experiment, and therefore the most interesting one from the astrochemical perspective.} Among other possible loss processes, thermal desorption should be negligible at 10~K, considering that much lighter species such as hydrogen atoms and \ce{H2} can also remain on the surface of solid \ce{CO} at this temperature.

After the deposition of C atoms was completed, a continuous decrease in C-atom intensity was clearly observed (Figure~\ref{fig:exp1}, panel~b). Decay profiles were recorded for various deposition times 100, 200, and 400~s, corresponding to the C-atom fluences of $2 \times 10^{12}$, $4 \times 10^{12}$, and $8 \times 10^{12}$~atoms~cm$^{-2}$. The decay profiles were well described by a single exponential function, indicating that the decay process follows first-order reaction kinetics. The exponential function is expressed as:

\begin{equation} \label{eq:exp2}
  \left[\textrm{C}\right]_{t} = \left[\textrm{C}\right]_{0}\left[ \left(1-b\right)\exp(-kt) +b \right]
\end{equation}
where $\left[\rm C\right]_{t}$ and $\left[ \rm C\right]_{0}$ represent the surface number density of C atoms at time $t$ and $t = 0$, respectively, $k$ is the decay rate, and $b$ is an asymptotic value indicating the fraction of C atoms that remain on the surface at infinite time. Due to the non-zero $b$ values ($b = 0.05$–$0.1$), the fitted curve is not a straight line even in the logarithmic scale plot. The $k$ values for deposition times of 100, 200, and 400 s were determined to be $(1.0 \pm 0.2) \times 10^{-2}$, $(7.7 \pm 1.2) \times 10^{-3}$, and $(6.3 \pm 1.2) \times 10^{-3}$ s$^{-1}$, respectively. These decay rates correspond to decay time constants ($1/k$) of $100 \pm 20$, $130 \pm 20$, and $160 \pm 30$ s, respectively. The errors represent statistical uncertainties across seven measurements.

We also analyzed the decay curves assuming the C-atom loss by diffusive recombination reaction (\ce{2C -> C2}). In this case, $\left[C\right]_{t}$ can be expressed as:
\begin{equation}
  \dfrac{\left[\rm C\right]_{0}}{\left[\rm C\right]_{t}} = 2 k_{\rm diff}\left[C\right]_{0}t+1
\end{equation}
where $k_{\mathrm{diff}}$ is the surface diffusion rate constant of C atoms on solid \ce{CO}. For C-atom deposition durations of 100 and 400~s, the values of $2 \cdot k_{\mathrm{diff}} \cdot [\mathrm{C}]_0$ were determined to be $0.012 \pm 0.006$ and $0.010 \pm 0.002$~s$^{-1}$, respectively. According to the experimental data presented in Figure~\ref{fig:exp1}a, the ratio of $[\mathrm{C}]_0$ values after 100 and 400~s depositions, $[\mathrm{C}]_0(400~\mathrm{s})/[\mathrm{C}]_0(100~\mathrm{s})$, is approximately 2.9. Under the assumption that the C-atom binding energy distribution is coverage-independent, the $k_{\mathrm{diff}}$ value should also be coverage-independent. Consequently, the $2 \cdot k_{\mathrm{diff}} \cdot [\mathrm{C}]_0$ value for the 400~s deposition should be about three times as large as that for the 100~s deposition if the C-atom decay is due to diffusive recombination. However, these values are consistent within statistical errors, indicating that diffusive recombination is not responsible for the observed C-atom decay. Therefore, any source of detectable C-atom loss must arise from processes other than C–C recombination. We propose that the only process responsible for the C-atom decay is the formation of \ce{^{3}CCO}, where the rate-limiting step is likely either the short-distance diffusion of C atoms to locate reactive \ce{CO} or the \ce{^{3}C + CO -> ^{3}CCO} reaction itself.

In order to get further insights into the decay process, its temperature dependence was studied. The decay profiles were measured at 5.5, 10, 15, and 20~K with 200~s deposition time. The C-atom decay rates at these temperatures were determined to be $(6.9 \pm 1.0) \times 10^{-3}$, $(7.7 \pm 1.2) \times 10^{-3}$, $(7.1 \pm 0.5) \times 10^{-3}$, and $(7.1 \pm 1.0) \times 10^{-3}$~s$^{-1}$, respectively. Hence, no temperature dependence was observed for the C-atom decay in the temperature range of 5.5–20~K, meaning that the C-atom decay proceeds by non-thermal mechanisms, such as quantum mechanical tunneling or barrierless reactions. A temperature-independent decay was also observed for C atoms on ASW \citep{tsuge_surface_2023}. In that case, the conversion from physisorbed to chemisorbed states by quantum mechanical tunneling was suggested as the most plausible process responsible for the decay, while the involvement of short-distance C-atom diffusion was excluded based on the experimentally determined activation energy for surface diffusion ($E_{\mathrm{sd}} = 1020$~K). Neither the activation energy for surface diffusion ($E_{\mathrm{sd}}$) nor for desorption ($E_{\mathrm{des}}$) has been reported for C atoms on solid \ce{CO}. \rev{Moreover,} as described in Section \ref{sec:discussion}, quantum chemical calculations suggest that the \ce{^{3}C + CO} reaction producing \ce{^{3}CCO} is barrierless. Therefore, we exclude quantum tunneling of C atoms as the primary reaction driver. Interestingly, quantum tunneling might still play an indirect role in the decay of the C atom signal in our experiment. As demonstrated by \rev{\citet{choudhury_condensed-phase_2022}}, \ce{CO} molecules undergo orientational isomerization (i.e., flipping of the CO dipole) at low temperatures via quantum tunneling on laboratory timescales. These timescales align remarkably well with our experimental observations, suggesting that the reactivity of C atoms may depend on the orientation of \ce{CO} molecules. Specifically, C–CO orientations are immediately reactive (as shown in Section \ref{sec:discussion}), while C–OC orientations are unreactive and require CO flipping via the aforementioned tunneling mechanism to enable reactions.

In addition to IR spectroscopy (shown in Section \ref{sec:experimental:pre}), we attempted to detect \ce{^{3}CCO} on solid \ce{CO} using the PSD-REMPI method, but without success. We employed PSD laser wavelengths of 532~nm or 355~nm, along with a REMPI laser wavelength of 450–453~nm \citep{tjossem_species_1985}. The non-detection of \ce{^{3}CCO} could be attributed to either (i) \ce{^{3}CCO} not being desorbed by the 532~nm or 355~nm PSD laser, or (ii) the lifetime of \ce{^{3}CCO} being too short, resulting in its surface number density falling below the detection limit due to rapid conversion, e.g., into \ce{C3O2} (see Sections \ref{sec:experimental:pre} and \ref{sec:discussion}). It is worth noting, however, that the reference for REMPI detection was obtained from gas-phase studies \citep{tjossem_species_1985}.

\subsection{Sequential deposition experiments} \label{sec:experimental:pre}

\begin{figure}
  \centering
  \includegraphics[width=\linewidth]{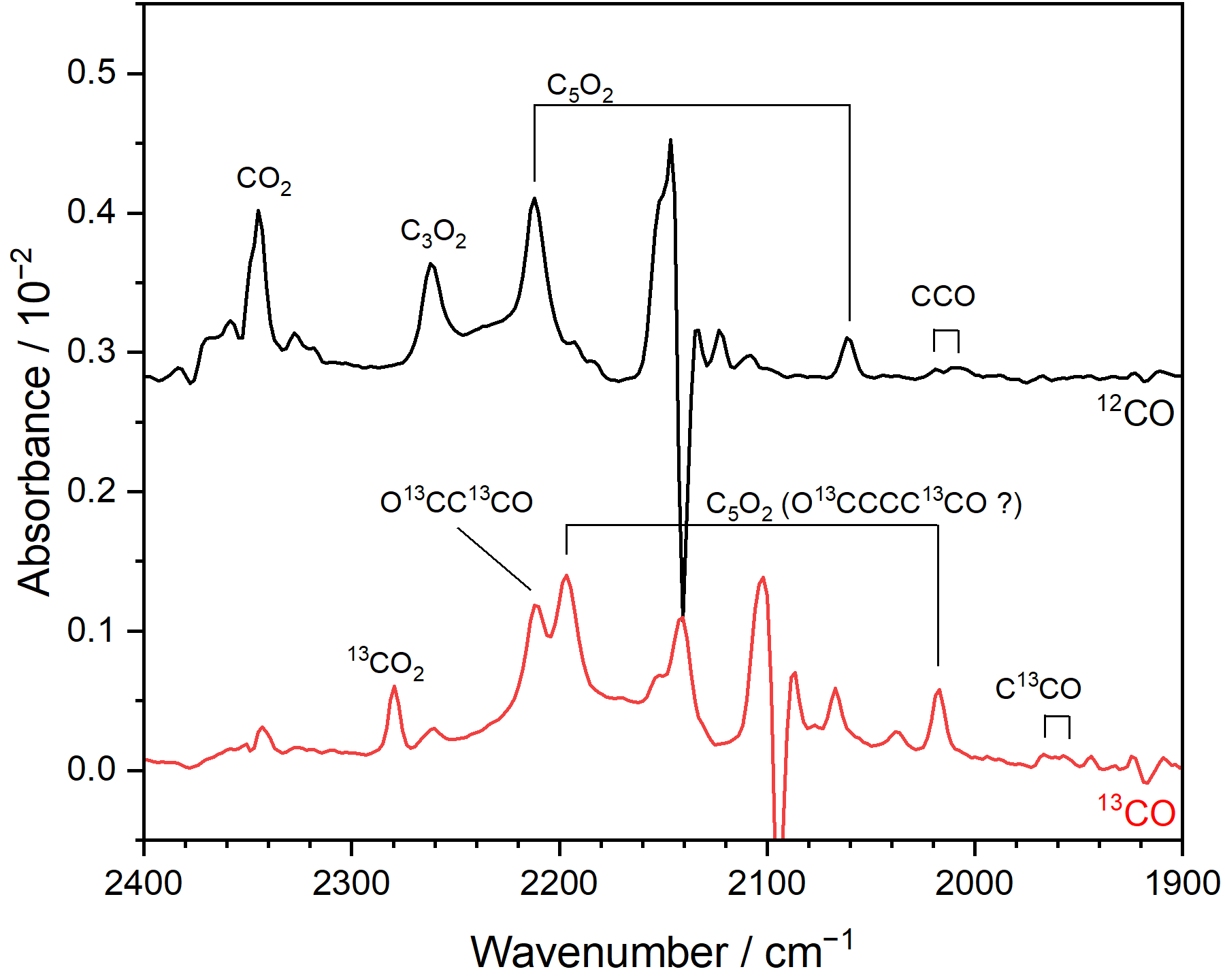}
  \caption{IR difference spectra measured after depositing C atoms on solid $^{12}$CO (upper trace) and on solid $^{13}$CO (bottom trace) at 10 K. The thickness of solid CO was about 10 MLs and the C-atom fluence was about 7$\times$10$^{13}$ atoms cm$^{-2}$.   Spectral assignments for \ce{CO2}, \ce{C3O2}, \ce{C5O2}, and CCO are presented. Isotopic composition of \ce{C5O2} in $^{13}$CO experiment is uncertain (see text).}
  \label{fig:exp2}
\end{figure}

Since the primary reaction product of the \ce{C + CO} reaction, \ce{^{3}CCO}, could not be identified using the PSD-REMPI method, we conducted measurements of IR spectra. These measurements involved both the sequential deposition of C atoms on solid \ce{CO} (this section) and the co-deposition of C atoms and \ce{CO} (Section~\ref{sec:experimental:co}). The sequential deposition closely resembles the PSD-REMPI experiments and is more representative of astrophysical conditions.

The upper trace of Figure~\ref{fig:exp2} shows the IR difference spectrum after 60 minutes of C-atom deposition on solids of \ce{^{12}CO} and \ce{^{13}CO} at 10~K, with C-atoms fluence of approximately $7 \times 10^{13}$~atoms~cm$^{-2}$.  Here, \ce{^{12}CO} and \ce{^{13}CO} refer to the normal CO sample (natural \ce{^{13}C}/\ce{^{12}C} ratio) and the \ce{^{13}C}-enriched sample (99.2 atom\%), respectively. In the spectrum obtained for solid \ce{^{12}CO} (upper trace), intense peaks were observed at 2062, 2212, 2262, and 2345~cm$^{-1}$, along with a weak feature near 2008~cm$^{-1}$. The peak at 2345~cm$^{-1}$ is clearly attributed to \ce{CO2}. Based on literature values \citep{gerakines_carbon_2001,jamieson_understanding_2006,trottier_carbonchain_2004,Boogert2015}, the peaks at 2062 and 2212~cm$^{-1}$ were assigned to \ce{C5O2}, the peak at 2262~cm$^{-1}$ to \ce{C3O2}, and the weak feature at 2008~cm$^{-1}$ to \ce{CCO}. The spectrum measured for solid \ce{^{13}CO} (Figure \ref{fig:exp2}, lower trace) also supports these assignments. The intensities of the peaks corresponding to products, except for \ce{CCO}, increased with deposition time, whereas the intensity of the \ce{CCO} peak reached its maximum after 20–30 minutes, corresponding to a C-atom fluence of $2$–$4 \times 10^{13}$~atoms~cm$^{-2}$, and remained nearly constant thereafter. Using calculated absorption coefficients (see Section~\ref{sec:exp:ir}), the following column densities were estimated for a C-atom fluence of approximately $7 \times 10^{13}$~atoms~cm$^{-2}$: \ce{CCO}, $5.3 \times 10^{12}$~cm$^{-2}$; \ce{C3O2}, $6.5 \times 10^{12}$~cm$^{-2}$; \ce{C5O2}, $2.0 \times 10^{12}$~cm$^{-2}$; and \ce{CO2}, $1.6 \times 10^{13}$~cm$^{-2}$. These column density estimates are uncertain because absorption coefficients calculated for gas-phase species were used, except for \ce{CO2} \citep{gerakines_first_2015}. For \ce{C3O2}, an experimental absorption coefficient has been reported ($1.3 \times 10^{-16}$~cm~molecule$^{-1}$), and using this value, the column density of \ce{C3O2} is calculated to be $2.4 \times 10^{13}$~cm$^{-2}$. As described below, the \ce{CO2} product does not incorporate deposited C atoms (i.e., the carbon in \ce{CO2} originates from \ce{CO} molecules). Therefore, the major species produced by consuming C atoms in the sequential deposition experiments is \ce{C3O2}. For comparison, we also performed C-atom deposition on crystalline \ce{CO} ($\alpha$–\ce{CO}) solid at 10~K, but no significant differences were observed between amorphous and crystalline \ce{CO} solids.

\begin{figure}
  \centering
  \includegraphics[width=\linewidth]{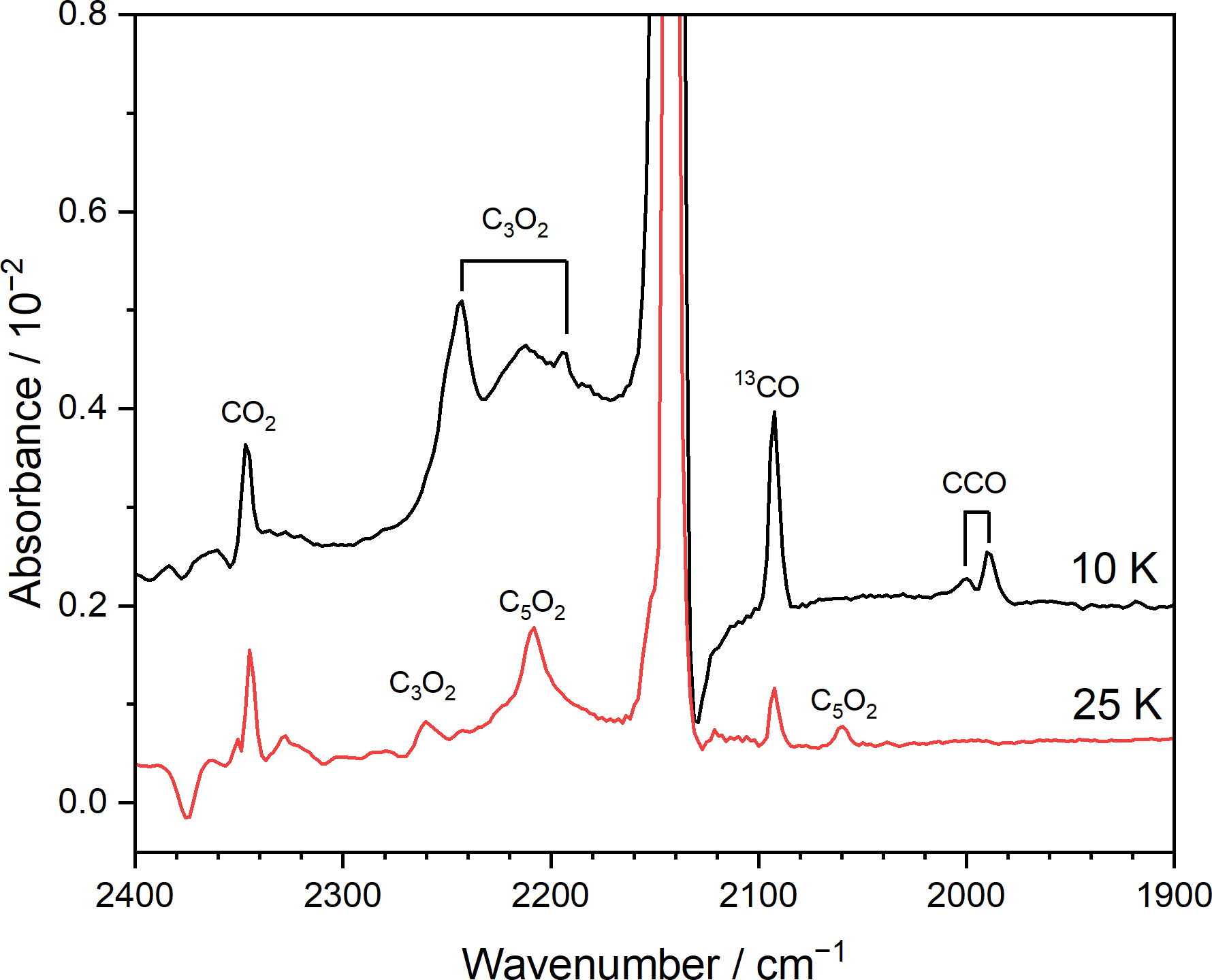}
  \caption{IR spectra measured in the C/CO co-deposition experiments. Deposition at 10 K  and deposition at 25 K (lower trace). The C/CO ratio was approximately 1/1000 and the C-atom fluence was about 5$\times$10$^{13}$ atoms cm$^{-2}$. Spectral assignments for \ce{CO2}, \ce{C3O2}, \ce{C5O2}, and CCO are presented. }
  \label{fig:exp3}
\end{figure}

\subsection{Co-deposition experiments} \label{sec:experimental:co}

Co-deposition experiments have been often conducted because reactant molecules are well mixed and have more chances to encounter during accumulation on the surface, leading to higher reaction yields. We performed co-deposition of C atoms and \ce{CO} for comparison, particularly with the study \rev{of \citet{fedoseev_hydrogenation_2022}}, which identified the formation of ketene (\ce{H2CCO}) in H/C/\ce{CO}/\ce{H2O} mixtures.

The upper trace of Figure~\ref{fig:exp3} shows the IR spectrum measured after co-deposition of C atoms and \ce{CO} (1/1000 dilution) at 10~K for 40 minutes, where the C-atom fluence was approximately $5 \times 10^{13}$~atoms~cm$^{-2}$. In the spectrum, a broad feature spanning from 2160–2300~cm$^{-1}$ is the phonon wing of solid \ce{CO}. In this spectral region, two peaks were identified at 2195 and 2243~cm$^{-1}$ and are assigned to \ce{C3O2}. The absence of any features near 2062~cm$^{-1}$, which was observed in the sequential deposition experiment (see Figure~\ref{fig:exp2}), indicates that \ce{C5O2} was absent in this sample. The IR features attributable to \ce{CCO} were observed at 1990 and 2000~cm$^{-1}$. In addition, \ce{CO2} was observed at 2347~cm$^{-1}$ similarly to the sequential deposition experiments. Column densities of these products were estimated for a C-atom fluence of about $5 \times 10^{13}$~atoms~cm$^{-2}$: \ce{CCO}, $7.3 \times 10^{13}$~cm$^{-2}$; \ce{C3O2}, $9.2 \times 10^{12}$~cm$^{-2}$; \ce{CO2}, $1.4 \times 10^{13}$~cm$^{-2}$. The \ce{CCO} yield exceeds the C-atom fluence most probably because the \ce{CCO} column density was overestimated. Nonetheless, \ce{CCO} seems to be the major product in the C/\ce{CO} co-deposition conditions. It should be noted that the \ce{CCO} formation is significantly enhanced when C and \ce{CO} were co-deposited probably because of rapid dissipation of reaction energy as high as 53.7~kcal~mol$^{-1}$ (Section~\ref{sec:discussion}). \rev{\citet{fedoseev_hydrogenation_2022}} reported that co-deposition of H/C/\ce{CO}/\ce{H2O} yielded ketene (\ce{H2CCO}) and \ce{C3O2} was absent in their spectra, indicating that hydrogenation of \ce{CCO} occurred readily so that \ce{CCO + CO -> C3O2} is hindered. Consequently, we suspect that ketene formation is enhanced only under laboratory conditions, where the \ce{CCO} product is readily thermalized and H atoms are abundant. Under molecular cloud conditions, typical H atom accretion rate to each grain is one H atom per day. Thus, the timescale for the H atom reaction is significantly longer than that of \ce{CCO + CO -> C3O2}, whose timescale is shorter than that of the energy dissipation process. Finally, we note that experimental conditions are quite different between this work and \rev{\citet{fedoseev_hydrogenation_2022}}; C-atom flux was about 25 times larger than ours, the C/\ce{CO} ratio of their work was approximately 1/2, and \ce{H2O} was the dominant component. 

The spectrum obtained after the deposition of a C/\ce{CO} = 1/1000 mixture at 25~K for 40 minutes is shown as the lower trace of Figure~\ref{fig:exp3}. The ratio of \ce{^{13}CO} intensity between 10 and 25~K indicates that the adsorption coefficient of \ce{CO} molecules is several times smaller at 25~K. The peaks at 2060 and 2208~cm$^{-1}$ are assigned to \ce{C5O2}, while the peak at 2260~cm$^{-1}$ is attributed to \ce{C3O2}. The difference in the peak position for \ce{C3O2} between 10~K and 25~K is due to the differing crystallinity of the solid \ce{CO}; it is amorphous at 10~K and crystalline at 25~K. The amounts of these products were proportional to the deposition time. Notably, \ce{CCO} was absent in the 25~K sample. Since the formation of \ce{C3O2} is most likely due to the reaction \ce{CCO + CO -> C3O2}, \ce{CCO} should still form at 25~K but is rapidly consumed to produce \ce{C3O2}. The formation mechanisms for \ce{C5O2} remain unclear. Stoichiometrically, three C atoms and two \ce{CO} molecules are required to produce \ce{C5O2}. The absence of \ce{C5O2} at 10~K and its presence at 25~K suggest that thermal processes play a role in \ce{C5O2} production. Given the low C-atom flux in the experiment ($2 \times 10^{10}$~atoms~cm$^{-2}$~s$^{-1}$), the simultaneous involvement of three C atoms at a single reaction site is unlikely without some enhanced surface mobility (e.g., at 10~K). The most plausible explanation involves the formation of \ce{C2} molecules through the \ce{C + C -> C2} non-thermal reaction in the experimental setup, followed by a barrierless addition to a neighboring \ce{CCO} molecule; i.e., \ce{C2 + CCO -> C4O}, \ce{C4O + CO -> C5O2}. This mechanism explains the enhanced abundance of \ce{C5O2} at 25~K. This is the mechanism simulated in our quantum chemical calculations (see Section\ref{sec:discussion}), where it is shown to be effective. An alternative reaction mechanism could involve \ce{C_{n}O} ($n = 2, 3$) species reacting with \ce{CO} to produce \ce{C_{n}} and \ce{CO2}: \ce{C_{n}O + CO -> C_{n} + CO2}. Although this reaction is expected to be endothermic, the heat of reaction at the formation of \ce{C_{n}O} species, i.e., \ce{C_{n-1} + CO -> C_{n}O}, could induce this reaction. Interestingly, however, this mechanism suggests that a single C atom can trigger the formation of several \ce{CO2} molecules, aligning with the relatively high yield of \ce{CO2} observed in both sequential and co-deposition experiments.

\section{Discussion} \label{sec:discussion}

The combined results from the three sets of experiments (Section \ref{sec:main:experimental}) reveal a rather complex chemical reaction network for C atoms adsorbed on solid \ce{CO} that requires explanation. In previous studies, \rev{\citet{fedoseev_hydrogenation_2022}} primarily reported the formation of \ce{H2CCO} in a co-deposition experiment involving C/\ce{CO}/H, a finding further supported by theoretical \rev{calculations from \citet{ferrero_formation_2023}}. However, in the more diluted environment explored in this work, the lack of accessible H atoms prevents the formation of \ce{H2CCO}. As a result, in the absence of such rapid hydrogen-driven chemistry, we observe the non-energetic formation of unsaturated carbon chains, a process previously suggested only in ices exposed to energetic irradiation \citep{jamieson_understanding_2006,forstel_pentacarbon_2016,trottier_carbonchain_2004,sie_key_2022,seperuelo_duarte_laboratory_2010,ciaravella_chemical_2016}, but shown to extend to cold, non-energetic chemistry. As demonstrated in Sections~\ref{sec:experimental:pre} and~\ref{sec:experimental:co}, our experiments reveal the synthesis of a diverse array of carbon chains, including \ce{CCO}, \ce{C3O2}, and \ce{C5O2}, as well as \ce{CO2}. From an astrochemical perspective, identifying the atomistic processes responsible for the formation of these species is crucial for understanding their contribution to the total carbon chain budget in solid \ce{CO}.

To help in identifying the specific atomistic processes that contibute to the formation of every possible carbon chain we have conducted a quantum chemical investigation trying to replicate the experiments in the laboratory. In the first place, we attempted to simulate the formation of the \ce{CCO}, \ce{C3O2}, and \ce{C5O2} carbon chains. For that, the following reactions are considered:
\begin{align}
  \ce{C + CO &-> CCO} \label{eq:chains:1} \\
  \ce{CCO + CO &-> C3O2} \label{eq:chains:2} \\
  \ce{C2 + CCO &-> C4O} \label{eq:chains:3} \\
  \ce{C4O + CO &-> C5O2} \label{eq:chains:4}.
\end{align}
From these reactions, reaction \ref{eq:chains:1} is the one directly transferable to the ISM, whereas reaction \ref{eq:chains:3} require a localized concentration of C atoms, which is less likely to occur in real astronomical ices (see the discussions below).  

For reaction \ref{eq:chains:1}, we initiated simulations by randomly placing a carbon atom in its electronic ground state (i.e., C($^{3}$P)) within a pre-optimized ice model consisting of 33 \ce{CO} molecules (see Section~\ref{sec:theory}). This procedure was repeated for three distinct binding sites to gather statistical insights. In all cases, the simulations consistently showed the direct formation of \ce{CCO} in its triplet state, aligning with prior theoretical predictions \citep{ferrero_formation_2023}. The force minimization process revealed that the reaction proceeds via a C–CO interaction. Experimentally, however, C–OC adsorption is also possible. This geometry does not directly lead to reaction and requires \ce{CO} dipole flipping \citep{choudhury_condensed-phase_2022}, which \rev{can} explain the observed time-dependent behavior in our experiments.

\begin{figure*}
  \centering
  \includegraphics[width=\linewidth]{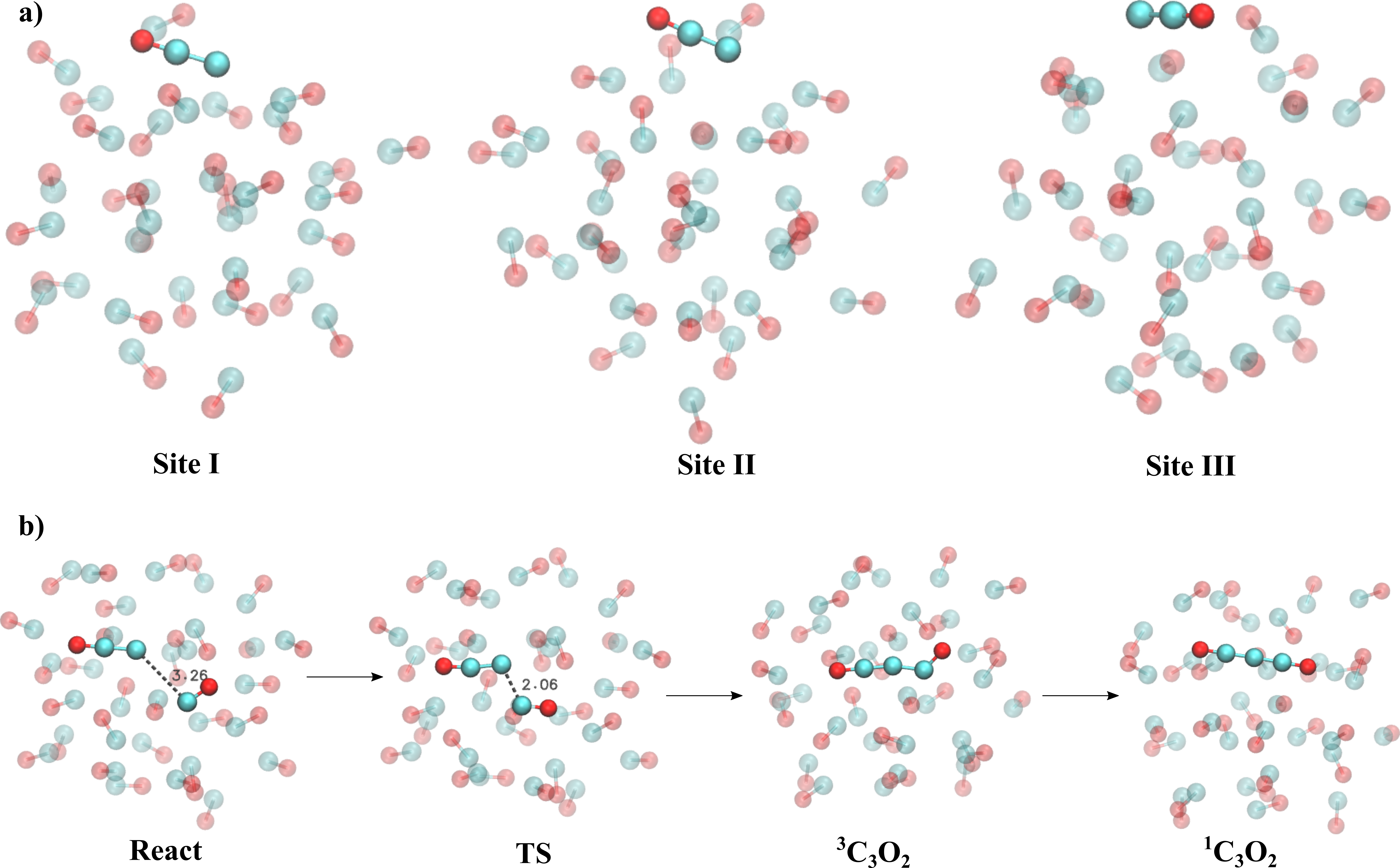}
  \caption{(a) Adsorption geometries of $^3$C on CO ice reflecting chemisorption and formation of the $^{3}$CCO complex on different binding sites at the DLPNO-CCSD(T)/CBS//MN15(D3BJ)/6-31+G(d,p) level. The individual values for the chemisorption energies are 54.0 \kcalmol for Site I, 54.0 \kcalmol for Site II and 53.0 \kcalmol~for Site III. (b) Scheme of the \ce{CCO + CO -> C3O2} reaction at Site I after $^3$C chemisorption.}
  \label{fig:theo1}
\end{figure*}

\begin{figure}
  \centering
  \includegraphics[width=\linewidth]{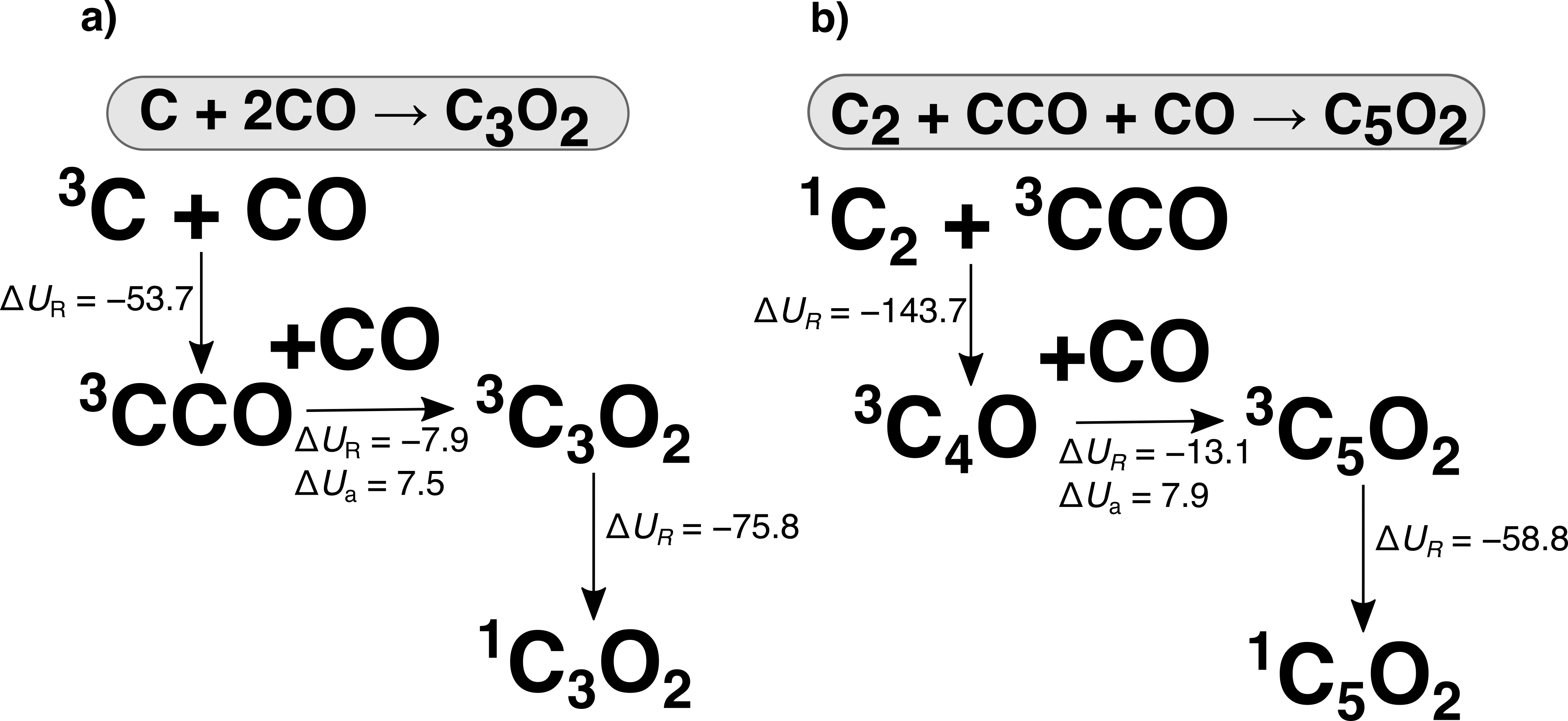}
  \caption{(a) Schematic view for the simulated reaction profile for the formation of \ce{C3O2} (b) Suggested scheme for the formation of \ce{C5O2} in our experiments. $\Delta U_{R}$ and $\Delta U_{a}$ represent the reaction energies and activation energies, if any, of the elementary steps showcased in the diagram. All energies are shown in \kcalmol.}
  \label{fig:theo3}
\end{figure}

The starting geometries for the subsequent calculations are shown in Figure~\ref{fig:theo1}, upper panel. The average adsorption energy, $\Delta U_{\mathrm{R}}$, was determined to be $-53.7 \pm 0.6$~kcal~mol$^{-1}$, which is approximately 4~kcal~mol$^{-1}$ higher than the reported gas-phase value \citep{papakondylis_electronic_2019}. This excess energy must correspond to the binding energy between \ce{^{3}CCO} and \ce{CO}. The chemisorption energy can be utilized in various ways, including transient diffusion on the \ce{CO} ice, desorption, \rev{surface reconstruction} or reaction with a neighboring \ce{CO} molecule. While the first \rev{three} processes are challenging to detect experimentally, the latter is straightforward to identify, as it corresponds to reaction~\ref{eq:chains:2}.  Using the three binding geometries shown in Figure~\ref{fig:theo1}, we examined reaction~\ref{eq:chains:2}, as depicted in the bottom panel of the figure for Site~I. Notably, the electronic ground state of \ce{C3O2} is a singlet, whereas \ce{CCO} is a triplet. To simulate reaction~\ref{eq:chains:2}, we first considered the formation of \ce{^{3}C3O2}, followed by an intersystem crossing that aligns with laboratory timescales, leading to \ce{^{1}C3O2}. The activation energy, $\Delta U_{\mathrm{a}}$, for reaction~\ref{eq:chains:2} in the triplet channel was found to be $7.5 \pm 0.7$~kcal~mol$^{-1}$, representing, on average, 13.9\% of the total available chemisorption energy. The $\Delta U_{\mathrm{R}}$ for the \ce{^{3}CCO + CO -> ^{3}C3O2} reaction was $-7.9 \pm 0.7$ ~kcal~mol$^{-1}$. Although limited information is available on chemical energy relaxation in solid \ce{CO}, recent studies on vibrational energy relaxation \citep{ferrari_vibrational_2024} suggest longer timescales than those observed in water ice \citep{molpeceres_reaction_2023,ferrero_where_2023}. This finding supports the argument for the rapid, non-thermal formation of \ce{^{3}C3O2} through reaction~\ref{eq:chains:2} in sequential deposition experiments. Moreover, it also explains why a small fraction of \ce{CCO} can be detected in experiments at 10~K, indicating a subset of binding sites with rapid energy dissipation. In co-deposition experiments, as shown in this work and by \citet{fedoseev_hydrogenation_2022}, \ce{CCO} is present because this dissipation occurs more quickly, especially in the presence of water molecules. \rev{An important aspect concerns the competition between \ce{C3O2} formation and the desorption of \ce{^{3}CCO} after its generation, given that the estimated adsorption energy is only about 4~kcal~mol$^{-1}$ (roughly 7.4\% of the overall reaction energy), see above. While chemical desorption (or diffusion) cannot be ruled out, its quantitative assessment lies beyond the scope of both our experiments and calculations. Nevertheless, recent studies on energy dissipation, albeit on \ce{H2O} ice \citep{molpeceres_reaction_2023}, demonstrate that the conversion of chemical energy from intramolecular modes into weak translational motion is highly inefficient, typically remaining below 7\%. In this context, the C=O stretching of \ce{^{3}CCO}, which corresponds to the reaction coordinate of \ce{^{3}CCO + CO -> ^{3}C3O2}, is far more likely to channel the excess energy towards reaction rather than desorption.} Finally, we did not explicitly simulate the triplet–singlet decay of \ce{C3O2}, but it is a plausible outcome within laboratory (and astronomical) timescales, considering the experimental evidence and the significant exothermicity of the process, calculated to be $-75.8 \pm 0.9$~kcal~mol$^{-1}$. A summary of the mechanism is shown in Figure \ref{fig:theo3}.

\begin{figure*}
  \centering
  \includegraphics[width=0.7\linewidth]{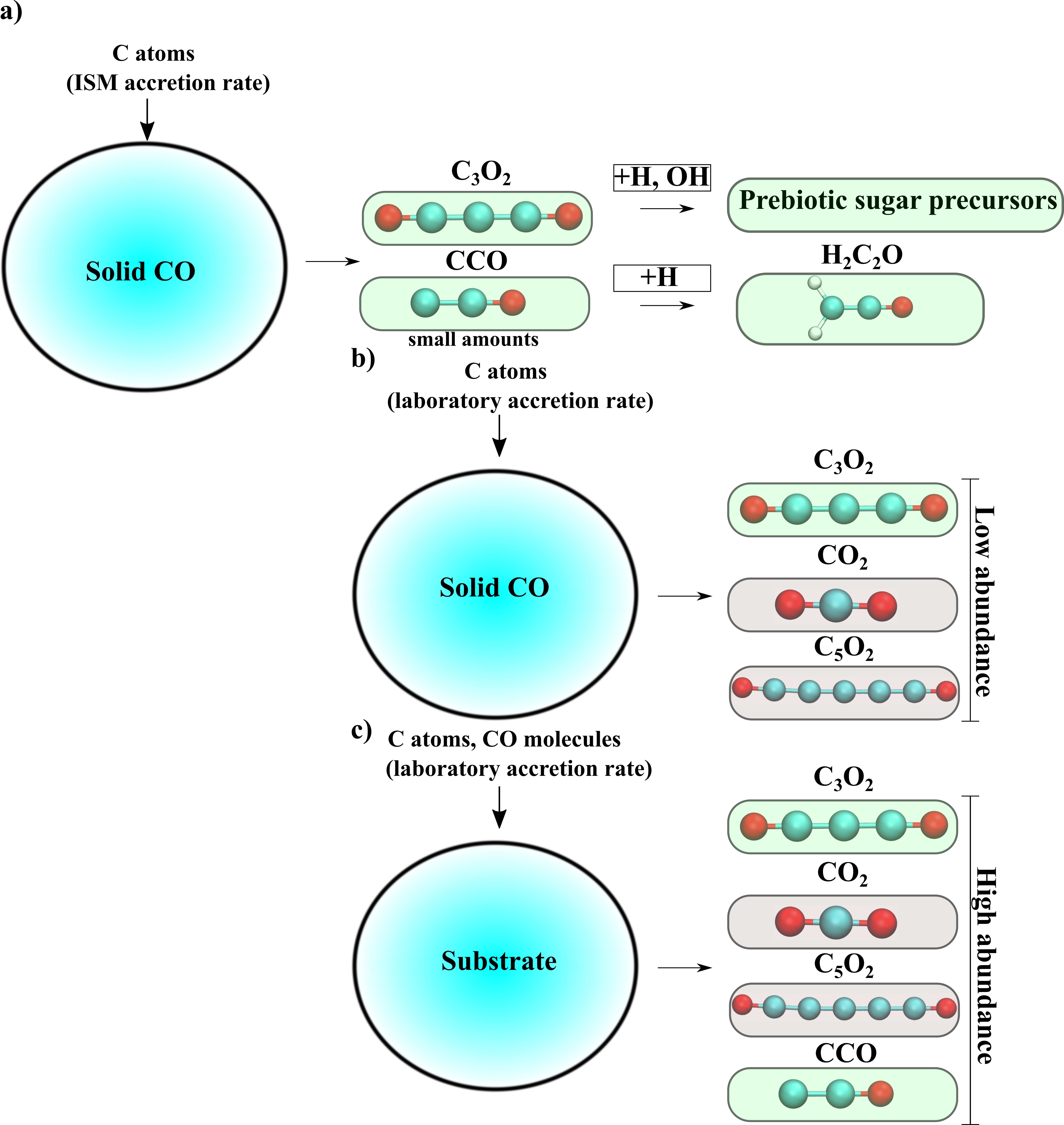}
  \caption{Schematic summary of our results. The figure illustrates the formation of oxygenated carbon chains, \ce{C3O2}, \ce{C5O2}, \ce{CCO}, and \ce{CO2}. Panel (a) represents the ideal situation in the ISM with a low accretion rate of carbon atoms impinging the ice, panel (b) shows the exemplification of the sequential deposition experiments at 10 K and panel (c) the co-deposition experiments at 10-25 K. Pale green boxes represent molecules that are compatible with ISM dilution conditions whereas gray boxes represent molecules that are less likely to occur in the ISM.}
  \label{fig:theo2}
\end{figure*}

The formation of \ce{C5O2} cannot be easily explained with a single $^{3}$C adsorption, because Reaction \ref{eq:chains:2} terminates the formation of the carbon chain. Likewise, succesive addition of $^{3}$C on a carbon chain is unlikely. Another, possibly more likely, option involves the reaction of \ce{C2} formed during the deposition with \ce{CCO} and subsequent reaction with a neighboring CO molecule. The mechanism for the formation of \ce{C2} might be due to direct association of two C atoms via a hot-atom mechanism. Regardless of the \ce{C2} formation mechanism, once formed it can react with \ce{CCO} to yield \ce{C4O} via reaction \ref{eq:chains:3}. Calculations for this reaction involve a strongly multireference electronic structure, in which the \ce{C2} molecule has two closely spaced electronic states, $^{1}\Sigma^{+}$ and $^{3}\Pi_{u}$ (as exemplified in the review by \rev{\citet{varandas_c_2018}}), therefore, our combined DFT and coupled cluster calculations should be viewed with caution, and only serve to prove a barrierless association between \ce{C2} and CCO. The calculations were performed placing a \ce{C2} molecule in the vicinity of a CCO radical ensuring a real electronic minimum in the triplet state through a wavefunction stability analysis, because the ground state of the \ce{C4O} molecule is a $^{3}\Sigma^{-}$ state \citep{kannari_ab_1994, rienstra-kiracofe_electron_2000,jamieson_understanding_2006}, and leaving the system evolve with a force minimization. In all cases, it was found that \ce{C2} evolves without barriers to \ce{C4O}. The depth of the \ce{C4O} well is $-$143.7 $\pm$ 0.8 kcal mol$^{-1}$. Therefore, when \ce{C2} is close to \ce{CCO} a spontaneous formation of \ce{$^{3}$C4O} takes place with an immense release of chemical energy. Once formed, the vast amount of reaction energy liberated in the association can be used to react with another neighboring CO molecule forming \ce{C5O2} (Reaction \ref{eq:chains:4}, and equivalent to the scheme shown in Figure \ref{fig:theo1} bottom panel) with only a 5.5\% of the association energy required for the reaction. Finally, the $\Delta U_{a}$ in the triplet channel is 7.9 $\pm$ 0.4 \kcalmol ($\Delta U_{R}$=$-$13.1 $\pm$ 1.1 \kcalmol) and the triplet-singlet gap is $-$58.8 $\pm$ 0.66 \kcalmol. Therefore, with this mechanism we explain the presence of \ce{C5O2} in our reaction mixture and only \ce{CO2} remains to be explained.  Regretfully, we could not explain the formation of \ce{CO2} in our simulations simply considering ground state chemistry, as we have done for all reactions in this section. It is well known that \ce{CO2} is readily formed by \ce{CO} in an excited state in heavily processed ices \citep{jamieson_understanding_2006, forstel_pentacarbon_2016, trottier_carbonchain_2004, 1996A&A...312..289G, sie_key_2022, seperuelo_duarte_laboratory_2010, ciaravella_chemical_2016,devine_spin-forbidden_2022}. Our results suggest that the formation of \ce{CO2} in our experiments might be a consequence of the high energy released in the formation of the carbon chains, forming C$_n$ and hence \ce{CO2} in the process. We consider the formation of C$_n$ and \ce{CO2} hardly transferable to the conditions of the ISM where the energy deposited by surface area is lower than in our experiments due to the lower accretion rate of C atoms. In summary, the formation of carbon chains observed in our experiments highlights the intricate chemical processes occurring on solid CO. However, the experimental conditions may not fully replicate the conditions prevalent in the ISM. Under ISM conditions, the formation of \ce{C3O2} appears plausible based on our calculations. However, the synthesis of longer carbon chains, such as \ce{C5O2}, or the production of \ce{CO2}, is less likely. From a physicochemical perspective, however, all the carbon chains identified in this study could theoretically form on nanosized icy grains. A schematic overview of the physicochemical and astrochemical pathways for carbon chain formation identified in this work is presented in Figure \ref{fig:theo2}.

Based on our experiments and quantum chemical calculations, it becomes evident that astrochemical reaction network models should include the chemisorption of C atoms on CO ice as well as that on \ce{H2O} ice. However, the resulting chemical pathways are likely sensitive to the local distribution of CO on the ice surface—whether CO molecules are randomly distributed or aggregated. In astrochemical models based on the rate-equation approach, which is the most widely used method in the astrochemistry community, adsorbed species on surfaces are implicitly assumed to be randomly and homogeneously distributed, which favors isolated encounters with CO and thus the formation of \ce{H2CCO}, depending on the surface coverage of CO. The surface coverage of CO would depend on the physical environment and time, ranging from approximately $<10\%$ to $\sim70\%$ \citep{taquet_multilayer_2014, molpeceres_enhanced_2024}. On the other hand, if CO molecules are aggregated, as suggested by \rev{\citet{Kouchi2021}}, the C-atom chemisorption on solid CO would lead to \ce{C3O2}, as demonstrated by this work. To assess such effects, the use of the microscopic kinetic Monte Carlo (kMC) approach, which can explicitly account for the spatial arrangement of adsorbed species on surfaces \rev{\citep{cuppen_kinetic_2013, garrod_three-dimensional_2013}}, would be more appropriate than the rate-equation approach, especially in the low to medium coverage of CO.

To conclude, the \ce{C3O2} molecule formed through this reaction chain in the ISM could serve as a precursor for further reactions under the unique conditions of interstellar ices. \rev{To our knowledge, \ce{C3O2} has not been firmly identified in the ISM. However, \citet{cartwright_jwst_2024} discussed their possible detection by JWST/NIRSpec. According to their interpretation, the features observed at 4.41 and 4.47 $\mu m$ might be due to either \ce{C3O2} (and higher order carbon chain oxides) or CN-bearing compounds known to exist in the irradiated ice containing carbon oxides and ammonia \citep{2007MmSAI..78..681S}.} 

\rev{Certainly, the conditions of our experiments and associated calculations are highly controlled, effectively restricting the chemical network under study to reactions involving carbon atoms, CO molecules, and their immediate products. In the interstellar medium, however, a far richer variety of reactive species is present, dramatically expanding the chemical complexity that can emerge following the readily formation of \ce{CCO} or \ce{C3O2}.} Of particular interest, as suggested by previous studies for the hydrogenation of \ce{C2O} \citep{fedoseev_hydrogenation_2022,ferrero_formation_2023}, is the hydrogenation of \ce{C3O2} facilitated by quantum tunneling. This process could lead to the formation of molecules such as \ce{H2C3O2}, \ce{H4C3O2}, \ce{H6C3O2}, or \ce{H8C3O2} \rev{(Figure \ref{fig:theo2})}. Beyond hydrogenation, reactions with other abundant radicals, particularly at the apolar-polar ice interface \citep{Boogert2015}, such as OH, NH$_2$, or CH$_3$, could yield even more complex molecules. These processes may significantly enhance the abundance of prebiotic molecules in interstellar ices. For instance, glycerol (\ce{C3H8O3}), glyceraldehyde (HOCH$_2$CH(OH)CHO), and glyceric acid (\ce{HOCH2CH(OH)COOH}), which are important prebiotic molecules identified in laboratory ice experiments \citep{kaiser_synthesis_2015, Fedoseev2017, wang_interstellar_2024}, could be synthesized more efficiently due to the increased availability of \ce{C3O2}. 

\section{Conclusion}

In this work, combining experimental and theoretical approaches, we have demonstrated for the first time the plausibility of forming complex carbon chains through the non-energetic reaction of atomic carbon with ice, complementing other experiments indicating that energetic input was needed. A key aspect of our findings and discussions is that not all experimental evidence is directly transferable to the chemistry of the ISM, where dilution conditions are far more extreme than in any laboratory experiment. After a careful analysis grounded in quantum chemical calculations, we conclude that the formation of \ce{C3O2} is the most likely reaction to occur in the ISM, while the formation of \ce{C5O2} and \ce{CO2} is likely an artifact of our experimental setup. We cannot rule out the stabilization of a small fraction of \ce{CCO} at low temperatures in the ISM. The facile synthesis of \ce{C3O2} in our experiments suggests that this molecule could serve as a precursor for the formation of more complex prebiotic sugars in \rev{real astrophysical environments}. We emphasize the importance of fundamental physical chemistry in proposing, understanding, and interpreting our experiments, along with the integration of experiments and theory to provide a comprehensive understanding of the molecular Universe.

\begin{acknowledgments}
This work was partially supported by JSPS KAKENHI (grant Nos. JP25K07364, JP25KJ133, JP24K00686, JP23H03982, JP22H00159, and JP17H06087). G.M acknowledges the support of the grant RYC2022-035442-I funded by MICIU/AEI/10.13039/501100011033 and ESF+. G.M. also received support from project 20245AT016 (Proyectos Intramurales CSIC). We acknowledge the computational resources provided by  the DRAGO computer cluster managed by SGAI-CSIC, and the Galician Supercomputing Center (CESGA). The supercomputer FinisTerrae III and its permanent data storage system have been funded by the Spanish Ministry of Science and Innovation, the Galician Government and the European Regional Development Fund (ERDF).
\end{acknowledgments}





%
\facilities{Cesga Supercomputer}

\software{\textsc{Gaussian16} \citep{g16}, \textsc{Orca6.0.1} \citep{Neese2012,Neese2020}
          }




\bibliography{sample701}{}
\bibliographystyle{aasjournalv7}



\end{document}